\documentclass[12pt]{article}
\newcommand{\be}[3]{\begin{equation}  \label{#1#2#3}}

\newcommand{\ee}{ \end{equation}}
\newcommand{\ba}{\begin{array}}
\newcommand{\ea}{\end{array}}
\newcommand{\p}{\partial}

\newcommand{\remark}[1]{}
\let\LARGE=\Large
\let\Large=\large
\begin{document}

%%%%%%%%%%%%%%%%%%%%%%%%%%%%%%%%%%%%%%%%%%%%%%%%%%%%%%%%

\thispagestyle{empty}
\rightline{CALT-68-2268}
\rightline{CITUSC/00-017}
\rightline{hep-th/0005185}

\vspace{8truemm}

\centerline{\bf \LARGE 
A non-singular infra-red flow 
%D=5 gauged supergravity and conformal
}

\bigskip

\centerline{\bf \LARGE 
from D=5 gauged supergravity
% infra-red fixed points in field theory 
}

\bigskip

\vspace{1truecm}

\centerline{\bf Klaus Behrndt\footnote{e-mail: 
 behrndt@theory.caltech.edu} }

\vspace{.5truecm}
\centerline{\em California Institute of Technology}
\centerline{\em Pasadena, CA 91125, USA}

\medskip

\centerline{\it CIT-USC Center For Theoretical Physics}
\centerline{\it University of Southern California}
\centerline{\it Los Angeles, CA 90089-2536, USA}

\vspace{1truecm}

%%%%%%%%%%%%%%%%%%%%%%%%%%%%%%%%%%%%%%%%%%%%%%%%%%%%%%%%

\begin{abstract}
We discuss domain wall solutions of 5-dimensional supergravity
corresponding to a cosine-superpotential, which is derived by a
gauging of the two Abelian isometries of the scalar coset
$SU(2,1)/U(2)$. We argue that this potential can be obtained from
M-theory compactification in the presence of G-fluxes and an M5-brane
instanton gas.  If we decouple the volume scalar of the internal
space, the superpotential allows for two extrema, which are either
ultra-violet or infra-red attractive.  Asymptotically we approach
therefore either the boundary or the Killing horizon of an
anti-deSitter space or flat spacetime for a vanishing cosmological
constant. If the volume scalar does not decouple, we obtain a run-away
potential corresponding to dilatonic domain walls, which always run
towards a vanishing cosmological constant.
\end{abstract}

%%%%%%%%%%%%%%%%%%%%%%%%%%%%%%%%%%%%%%%%%%%%%%%%%%%%%%%%%%%%%%%%

\newpage

%%%%%%%%%%%%%%%%%%%%%%%%%%%%%%%%%%%%%%%%%%%%%%%%%%%%%%%%%%%%%%%%%

\section{Introduction}

%%%%%%%%%%%%%%%%%%%%%%%%%%%%%%%%%%%%%%%%%%%%%%%%%%%%%%%

In the past years many aspects of domain wall (DW) solutions of
5-dimensional supergravity have been discussed and one of the most
interesting application concerns a supergravity description of the
infra-red (IR) physics of 4-dimensional field theories.  In particular
it is possible to obtain a supergravity description of the
renormalization group (RG) flow \cite{360,040,120} towards
non-trivial IR fixed points, which can be conformal or
non-conformal. Most progress has been made for non-conformal IR fixed
points \cite{330, 310}, where the scalars flow towards a
singularity in the superpotential.  Many superpotentials coming from
string theory compactifications have poles, which are IR attractive
\cite{150} and imply a singularity in the supergravity solution
\cite{010, 230}. In special cases the singularity indicates the
appearance of a Coulomb branch \cite{320} or can be resolved as
discussed in \cite{300}.  On the other hand less is known about
conformal IR fixed points.  Especially for supersymmetric RG-flows the
construction of the corresponding supergravity solutions prove
difficult \cite{090} and the only known example that we are aware
of has been discussed in \cite{040,111}.  Moreover, potentials with
IR-attractive fixed points are essential for a string- or M-theory
embedding of the Randall-Sundrum (RS) scenario \cite{060}. Only if
such potentials exist a thin-wall approximation is justified and we
can approximate the scalars as constants given by their fixed-point
values.

In this letter we discuss domain wall solutions,
where the superpotential in the simplest case depend on a single scalar
$\theta$ and has the form
\be002
W = 2 \, (a + b \cos\theta) \ .
\ee
Potentials of this type are generated naturally by an
instanton/monopole condensation and have an old history in gauge theory 
in the discussion of confinement \cite{673} and for domain
walls \cite{674}; for a recent discussion see also \cite{677}.

As we show in the next section, this superpotential can be obtained
from a model where the scalar fields parameterize the coset
$SU(2,1)/U(2)$ and a linear combination of both Abelian isometries is
gauged. Depending on the constant parameters $a$ and $b$ it yields
ultra-violet (UV) as well as IR attractive fixed points and the
extrema of $W$ give a (negative) cosmological constant yielding an anti
deSitter spacetime.  In the special case of $a=\pm b$, the cosmological
constant vanishes at one extrema and we obtain flat spacetime.

In section three we derive explicit domain wall solutions
interpolating between the (different) extrema of $W$.  Depending the
choice for $a$ and $b$, these solutions are not only interesting from
the RG-flow point of view, but provide also a realization of the  RS
scenario; the no-go statements \cite{090} do not apply for this
model.

Embedding this model into $N$=2, $D$=5 gauged supergravity
\cite{080}, the scalars parameterizing the $SU(2,1)/U(2)$
coset build up the universal hypermultiplet. As we argue in the last
section, from the M-theory perspective the potential can be understood
by a superposition of non-trivial G-fluxes and an M5-brane instanton
gas.

%%%%%%%%%%%%%%%%%%%%%%%%%%%%%%%%%%%%%%%%%%%%%%%%%%%%%

\section{Abelian gauging of the $SU(2,1)/U(2)$ coset}

%%%%%%%%%%%%%%%%%%%%%%%%%%%%%%%%%%%%%%%%%%%%%%%%%%%%%

In 5-d supergravity obtained by string or M-theory compactification,
the scalars parameterize a coset space and the only potential
consistent with supersymmetry comes from gauging of isometries of this
coset. Since we are interested in flat domain walls, we can omit the
gauge fields and the bosonic part of the action reads
\be020
S = \int d^5 x\ \sqrt{-g}\, \Big[\, {\frac{1}{2}} \, R - V(\Phi) - 
{\frac{1}{2}} g_{MN} \partial_{\mu} \Phi^M \partial^\mu \Phi^N
\Big] \ 
\ee
where the scalars $\Phi^M$ are real and parameterize a space
with the metric $g_{MN}$. Following general arguments \cite{070},
the potential can be expressed in terms of a superpotential $W$ as
\be030
V \ = 6 \, \Big( \, {3 \over 4}\, g^{MN} \partial_M W \partial_N W - W^2
\,
\Big)\ .
\ee
Before we can discuss BPS domain wall solutions, we have to derive the
superpotential.  We consider the coset $SU(2,1)/U(2)$ and the
metric can be derived from the K\"ahler potential (in \cite{007}
more general models will be discussed)
\be080
K = - \log(1- |z_1|^2 - |z_2|^2 )\ , \qquad |z_1|^2 +|z_2|^2 < 1\ .
\ee
There is another commonly used parameterization of this coset, which
makes the quaternionic nature of this manifold manifest and
corresponds to the K\"ahler potential $K = - \log({S+\bar S \over 2} -
C\bar C)$ \cite{251}, where the complex scalars $S,\, C$ are known to
enter the universal hypermultiplet of $N=2$ supergravity. In the
ungauged case both parameterizations are equivalent, 
% are related by a symplectic transformation and they 
% but gauging does not preserve the duality group and, in fact, 
but after gauging the resulting superpotentials differ
significantly\footnote{A similar effect is also known for $N=4$, $D=4$
supergravity, where the $SO(4)$ and $SU(2) \times SU(2)$ formulation
are duality equivalent in the ungauged case, but differ after gauging.
The first case has an AdS vacuum, the other not, see \cite{623}.
Note, gauging does not preserve the duality symmetry.}; in
the quaternionic formulation we could not find a superpotential with
at least two extrema. This may be related to the fact, that
it is non-trivial to introduce global quaternionic coordinates
on a curved space, which would be necessary for the discussion of
domain walls; but this issue deserves further clarification
\cite{007}. 
% Nevertheless, the interpretation of the four real
% scalars to build up the universal hypermultiplet is still valid. 
We do not want to go into details here about the relation of the two
models and will instead come back to our parameterization given by the
K\"ahler potential (\ref{080}). So, we will treat this 4-d scalar
manifold not as a quaternionic but as a special K\"ahler space and
follow the formalism developed in the literatur \cite{200}.  
A disadvantage is however, that it is not clear whether this procedure
yields necessarily supersymmetric solution with four unbroken
supercharges (because in the case at hand $z_1$ and $z_2$ are in the
same multiplet). A consistency check is, that the scaling dimensions
coming from the sugra scalars fit into known representations 
in the dual field theory. We come back to this point below.

The two phase rotations of $z_1$ and $z_2$ are two Abelian isometries
corresponding to the Killing vectors 
\be100
k_1 = z_1 \partial_{z_1} - \bar z_1 \partial_{\bar z_1}
\qquad , \qquad
k_2 = z_2 \partial_{z_2} - \bar z_2 \partial_{\bar z_2} \ .
\ee
It appears more convenient to introduce polar coordinates, as in
\cite{160}, given by
\be110
z_1 = r \, (\cos\theta/2) \, e^{i (\psi + \varphi)/2} \qquad , \qquad
z_2 = r \, (\sin\theta/2) \, e^{i (\psi - \varphi)/2}
\ee
and the scalar field metric $ds^2 = \partial_{\bar z^i}\partial_{z^j}
K\, d\bar z^i dz^j$ becomes
\be120
ds^2 = {dr^2 \over (1 -r^2)^2} + 
{r^2 \over 4 (1 - r^2)^2} \Big(d\psi + \cos\theta \, d\varphi \Big)^2 +
{r^2 \over 4(1 -r^2)}\Big(d\theta^2 + \sin^2\theta \, d\varphi^2 \Big) \ .
\ee
The K\"ahler form is exact and can be written as 
\be140
K_{uv} \,  dq^u \wedge dq^v
 = d\hat \omega \ , \qquad  {\rm where} \quad
\hat \omega = {i \over 2} {r^2 \over 1 - r^2 } \tau_3
\ , \quad \tau_3 = d\psi + \cos\theta \, d\varphi
\ee
with $q^u = (r, \theta, \varphi, \psi)$. Next, we gauge a general linear
combination of the two Killing vectors $k_1, k_2$ with some constants $a,b$ 
\be170 
k = a\, (k_{1} + k_{2})/2 + b\, (k_{1} - k_{2})/2 =
- 4i \, ( a \, \partial_{\psi} + b \, \partial_{\varphi}) 
\ee 
and the scalar derivative becomes $D_{\mu}
q^u = \partial_{\mu}q^u + k^u A_{\mu}$, where $A_{\mu}$ is the
graviphoton (we do not consider vector multiplets).  The
superpotential is given by the Killing prepotential $W = P$ which is
defined by the K\"ahler 2-from \cite{200}
\be160
K_{vu} k^u = - \partial_v P 
\ee
and becomes
\be180
W = P = {r^2 \over 2(1 - r^2 )} \, \Big(a + b \, \cos\theta \Big)  \ .
\ee
In addition, the Killing spinor equations get corrected by $\sim W \,
\Gamma_{\mu} \epsilon$ for the gravitino variation and $\sim i k^u
\epsilon$ for the hyperino variation. There are numerous different
conventions yielding different factors in these variations, but they
can be fixed by the condition, that the vacuum is given by extrema
of $W$ and that the difference of the extrema on each side of the wall
gives the energy stored by the wall. Moreover, solutions of the BPS
equations solve also the equations of motion for our Lagrangian
(\ref{020}); we come back to these equations in the next section.

%%%%%%%%%%%%%%%%%%%%%%%%%%%%%%%%%%%%%%%%%%%%%%%%%%%%%%%%%%

\section{BPS domain wall solution}

%%%%%%%%%%%%%%%%%%%%%%%%%%%%%%%%%%%%%%%%%%%%%%%%%%%%%%%%%%

In supergravity one refers to domain walls as kink solution
interpolating between different extrema of the potential.  As an
Ansatz for the metric which is adapted to the RG flow interpreation
and preserves 4-d Poincare invariance one can use
\be010
ds^2 = \mu^2 \Big( -dt^2 + d\vec x^2 \Big) + { d\mu^2 \over 
\Big(\mu \widehat W(\mu)\Big)^2}\ ,
\ee
where the fifth coordinate $\mu$ will be identified with an energy
scale in the dual 4-d field theory. In these coordinates the UV region
(= large length scale in supergravity) corresponds to $\mu \rightarrow
\infty$, whereas the IR is approached for $\mu \rightarrow
0$. 
% $\widehat W(\mu)^2$ becomes the cosmological constant at the
% conformal fixed points.  
For our purpose we are only interested in the
dependence of the scalar fields on the fifth coordinate or, in terms
of the dual field theory, on the energy scale $\mu$. 
Using this ansatz for the metric the first order
equations that solve the equations of motion are \cite{020}
\be042
W= \pm \widehat W \qquad , \qquad \beta^M \equiv \mu {d \over d\mu} \Phi^M 
= -3 \, g^{MN} \partial_N \log W  \ .
\ee
Using the projector $\Gamma^5 \epsilon \ = \pm \epsilon$ the first
equation is equivalent to the gravitino and the second to the hyperino
variation. Oviously, supersymmetric extrema of $V$ occur at
$\partial_M W =0$ where also $\beta^M=0$ holds. For $W|_{\partial
W=0}\neq 0$ one has an AdS space with $W$ being the cosmological
constant while $W|_{\p W=0} =0$ corresponds to a flat space
time. {From} the field theory point of view, $\beta^M$ is a natural
candidate for the $\beta$-functions of the couplings related to the
supergravity scalar fields. For BPS solutions these functions
determine the holographic RG flow \cite{120,020}.

The nature of the fixed point is determined by the derivatives of the
$\beta$-functions\footnote{Here we assume that the fixed point is non
singular i.e. the metric non-degenerate.}
\be779
\p_N \beta^M|_{\beta=0} \ = \ -3\,g^{MK}\,\frac{\p_N\p_K
W}{W}\, \Big|_{\beta=0}~.
\ee
In the case that all scalars are in vector multiplets one finds $\p_M
\beta^N|_{\beta=0} = - 2 \delta_M^N$ and all fixed points are
necessarily UV attractive \cite{020,090}. Consequently, these
models cannot describe a smooth RG flow or give a smooth RS scenario,
where each side of the wall has to be IR attractive.  This situation
is very typical for many potentials coming from string or M-theory
compactification, which do not allow for ``good'' domain walls with at
least two smoothly connected extrema. The situation in four dimensions
is better \cite{030} (for a review see \cite{100}), but 5-d
supergravity is more restrictive. In fact, generically one has no
isolated extrema and instead a ``run-away'' potential giving rise to
dilatonic domain walls. This behaviour is caused by the scalar field
parameterizing the volume of the internal space. Whenever this scalar
is dynamical it runs either to zero or infinite volume.  There is no
mechanism known to stabilize the volume at a finite value, at least
not in a supersymmetric way. This happens also in our case, where the
radial part allows only for a single extrema where the superpotential
vanishes and the spacetime becomes flat.  On the other hand the volume
scalar can become non-dynamical.  For example, quantum corrections in
string or $M$-theory can provide a natural lower bound on the volume
\cite{170}, which cut-off the radial flow, see \cite{140}.  Or,
adopting the procedure discussed for 4-d vacua \cite{370}, we can
consider an infinite volume limit (non-compact Calabi-Yau), which
becomes equivalent to treat the $r$-coordinate as non-dynamical (i.e.\
constant). In doing this we can normalize $g_{\theta\theta} =1$
and consider first the model for the superpotential
\be200
W= 2 \, (a +  b \, \cos\theta)  \ .
\ee
At the end we will also
comment on the solution coming from a running radial coordinate.
Coming back to our analysis from before, this superpotential has
the properties we are looking for. It has two extrema where
$\cos\theta = \pm 1$ with
\be210
\partial_{\theta} \beta^{\theta} = 3b \, { b + a \cos\theta
\over (a + b \cos\theta)^2} \qquad , \qquad
\Delta = \partial_{\theta} \beta^{\theta}\Big|_{\beta=0} = 
{ 3 b \over b \pm a}
\ee
where $\Delta$ corresponds to the scaling dimension of the
corresponding field theory operator.  Therefore, for $a>b>0$ it
describes a flow from the UV at $\cos\theta = -1$ towards the IR at
$\cos\theta =1$. If $b>a \geq 0$ we have two IR fixed points and
similarly for $0 > a > b$ one finds on each side of the wall an UV
fixed point. Finally for $a=b > 0$ it is a domain wall between an IR
point at $\cos\theta = 1$ and flat spacetime at $\cos\theta =-1$.
Thus, this simple model describes all known types of supersymmetric
domain wall solutions. If both sides have the same type of fixed
points, $W$ has to change its sign and the solution cannot be
interpreted as RG-flow, see also \cite{270}. The pole in the
$\beta$-functions indicates a first order phase transition.

As we mentioned earlier, in order to trust the embedding of our model
into $N=2$ supergravity, we have to ensure, that the scaling
dimensions fit into short representations of superfields of the dual
field theory, see e.g.\ \cite{040}, and there are some interesting
cases. If we gauge e.g., only the $\partial_{\varphi}$ isometry we
obtain $a=0$ and $\Delta =3$; if we gauge only $k_1$ or $k_2$ we have
$a = \pm b$ with $\Delta = {3 \over 2}$ or if we turn off the
$\partial_{\varphi}$ gauging there are no running hyper scalars and we
obtain a special case of the model discuss in \cite{020} with $\Delta
= 2$ for the vector scalars.

Next, let us construct the explicit domain wall solution.
% we have to solve the BPS equations, 
% which are equivalent to the vanishing of the
% gravitino and of hyperino variation. 
The coordinate system used before was adapted for the discussion of
the RG flow, but in order to find an explicit solution we write the
metric as
\be230
ds^2 = e^{2 A(z)} \Big(-dt^2 + d\vec x^2 \Big) + dz^2 \ .
\ee
For these coordinates the BPS equations become \cite{010}
\be240
\partial_z A = W \ , \qquad \partial_z \theta = - 3 \, 
g^{\theta\theta} \partial_{\theta} W  \ .
\ee
and inserting the superpotential (\ref{200}) and $g_{\theta\theta}
=1$, we find as solution
\be250
e^{2 A} = e^{4az} \, (\cosh6bz)^{- 2/3} \ , \qquad
\cos\theta = - \tanh 6bz \ .
\ee
Approaching the two AdS vacua at $z = \pm \infty$, the warp factor
becomes $e^{2A} \simeq e^{4(\pm a - b) |z|}$ and as mentioned before
if $a>b>0$ we have an UV fixed point at $z = + \infty$ and an IR fixed
point on the other side. If $b>a>0$ there are IR fixed point on both
sides, which becomes $Z_2$ symmetric if $a=0$.  In this case, $e^{2A}$
is exponentially decreasing on both sides yielding a localization of
gravity on the wall and our model describes a Randall-Sundrum scenario
\cite{060}.  Having this thick wall, one can also consider a thin-wall
approximation ($b \rightarrow \infty$), where the scalar becomes
constant and the spacetime is everywhere AdS, up to the discontinuity
at $z=0$\footnote{Supersymmetric versions of the Randall-Sundrum
geometry have also been discussed in refs.  \cite{271}.}.  Another
interesting example is $a=b$, where we find
\be252
e^{-2A} = ( 1 + e^{- 12 a z})^{2/3}
\ee
and the domain wall represents a flow from an IR fixed point at $z = -
\infty$ to flat spacetime at $z= + \infty$.

Finally let us comment on the case of a running radius $r$, i.e.\ we
consider the complete superpotential (\ref{180}). In the
$\theta$-equation in (\ref{240}), the radial part drops out and we
obtain the same solution $\cos\theta = - \tanh 6bz$ and in addition we
have the equation for the radius
\be260
\partial_z r = -3 g^{rr} \partial_r W = -3 r (a + b \cos\theta) 
= -3 (a - b \tanh 6bz)
\ee
which is solved by $r^2 = e^{-6az} \, \cosh6bz$ for $a>b$ (keeping in
mind, that $0 \leq r < 1$). Therefore, $z=0$ corresponds to the singular
point $r=1$, where the superpotential and the potential $V$ have a pole
and the warp factor in the metric has a zero: $e^{2A} \sim
z^{1/6}$ for $z\simeq 0$. 
%  and this model gives a further example for a
% violation of the bound discussed in \cite{141}. 
Towards larger values of $z$ the warp factor increases and for $z= +
\infty$ corresponding to $r=0$ the potentials vanish and the spacetime
becomes flat. Hence, taking the radial part into account, there are
neither UV nor IR fixed points related to an AdS spacetime. Notice
that using our definition from before, the $\beta$-function $\beta^r =
-6 ({1 \over r} - r)$ vanishes only at the singularity $r=1$, which is
IR attractive, as the cases discussed in \cite{150}.

%%%%%%%%%%%%%%%%%%%%%%%%%%%%%%%%%%%%%%%%%%%%%%%%%%

\section{Some comments about the M-theory embedding}

%%%%%%%%%%%%%%%%%%%%%%%%%%%%%%%%%%%%%%%%%%%%%%%%%%%

Perhaps the easiest way to understand the solution comes from
$M$-theory, where the four scalars of the universal hypermultiplet can
be understood as follows. One scalar parameterizes the volume of the
internal space, one axion comes from the dualization of the 5-d 4-form
field strength, and two scalars are related to membranes wrapping
$\Omega_{0,3}$ or $\Omega_{3,0}$ cycle.  In our parameterization we
gauged the Killing vector $k \sim a\, \partial_{\psi} + b \,
\partial_{\varphi}$. The case $b=0$ reproduces the results derived in
\cite{190, 140} and therefore corresponds to turning on 4-form fluxes
in the $M$-theory compactification, which is equivalent to a
non-trivial M5-brane background.  On the other hand gauging the
$\partial_{\varphi}$ isometry gives a mass term for the $\theta$
axion. In fact for this deformation (setting $a=0$) the supergravity
potential can be written as (normalizing $g_{\theta\theta} = 3/4$ and
rescaling $\theta \rightarrow {2 \, \theta \over \sqrt{3}}$): $V = -6b
\, \cos {4 \theta \over \sqrt{3}}$ and we obtain a sine-Gordon model
coupled to gravity.  This form suggests, that it can be understood as
coming from a non-trivial instanton background.  M-theory compactified
to 5 dimension may give a topological term $\int d^5 x \, \theta \,
dG$. If there are no sources present, the Bianchi identity implies $dG
= 0 $ and such a term vanishes. If on the other hand magnetic sources
for $G$ exist in 5 dimensions, then $dG $ is non-vanishing, and this
term is the 5-d analogue of the familiar universal axionic coupling
$\int d^4x \, \psi \, tr ( F \wedge \tilde{F})$ in 4 dimensions. An
obvious candidate for magnetic sources of $G$ are M5-branes. However
in order to generate a cosine potential for $\theta$ such sources must
be pointlike in 5 dimensions (like the instanton density $tr (F \wedge
F )$ in 4 dimensions ) and therefore correspond to Euclidean
M5-branes wrapping the whole CY manifold, in contrast to the 5-branes
associated with the gauging of the $\partial_{\psi} $ isometry, which
wrap a holomorphic 2-cycle and become 3-branes upon
compactification. As in 4 dimensions, see \cite{676}, we expect that
the sum over an M5-brane instanton gas will reproduce the cosine
potential.
% and if this picture is true, there are two competing effects
% producing the RG flow: G-fluxes and M-brane instantons. 
Clearly, this interpretation deserves a more detailed investigation.

\bigskip

{\bf Acknowledgements}

The work is supported by a Heisenberg grant of the DFG. I am grateful
to S.\ Ferrara, S.\ Gukov, C.\ Pope and A.\ Zaffaroni for discussions
and comments. I would also like the thank C.\ Herrmann, J.\ Louis and
S.\ Thomas for collaboration on part of this project.

%%%%%%%%%%%%%%%%%%%%%%%%%%%%%%%%%%%%%%%%%%%%%%%%%%%%%%%%%%%%%%

% ---- Bibliography ----

% \nocite{*}                   %this uses *everything* in the .bib file
% \bibliography{mar00}          %or whatever your .bib file is
% \bibliographystyle{utphys}   %if you use utphys.bst

\providecommand{\href}[2]{#2}\begingroup\raggedright\endgroup

\end{document}